\documentclass{sig-alternate}

\usepackage{url}
\usepackage{amsmath}
\newtheorem{definition}{Definition}

\newtheorem{theorem}{Theorem}

\usepackage{changepage}
\usepackage{algorithm}
\usepackage{algorithmic}
\usepackage{multirow}
\usepackage{amsmath}
\usepackage{xcolor}
\usepackage{color}
\usepackage{times}
\usepackage{mathptmx}
\usepackage{mathtools}

\newcommand{\QED}{\mbox{\rule[0pt]{1.0ex}{1.0ex}}}
\def\boxend{\hspace*{\fill} $\QED$}

\newcommand{\nop}[1]{}

\newcommand{\tabincell}[2]{\begin{tabular}{@{}#1@{}}#2\end{tabular}}

\newtheorem{example}{Example}
\newtheorem{assumption}{Assumption}

\begin{document}
\begin{sloppy}
%

\title{In-Network Neighborhood-Based Node Similarity Measure: A Unified Parametric Model}

\numberofauthors{2}
\author{
\alignauthor
Yu Yang\\
       \affaddr{School of Computing Science}\\
       \affaddr{Simon Fraser University}\\
       \affaddr{Burnaby, BC, Canada}\\
       \email{yya119@sfu.ca}
\and
\alignauthor
Jian Pei\\
       \affaddr{School of Computing Science}\\
       \affaddr{Simon Fraser University}\\
       \affaddr{Burnaby, BC, Canada}\\
       \email{jpei@cs.sfu.ca}
}

\maketitle
\begin{abstract}

In many applications, we need to measure similarity between nodes in a large network based on features of their neighborhoods.  Although in-network node similarity based on proximity has been well investigated, surprisingly, measuring in-network node similarity based on neighborhoods remains a largely untouched problem in literature.  One grand challenge is that in different applications we may need different measurements that manifest different meanings of similarity.  In this paper, we investigate the problem in a principled and systematic manner.  We develop a unified parametric model and a series of four instance measures.  Those instance similarity measures not only address a spectrum of various meanings of similarity, but also present a series of tradeoffs between computational cost and strictness of matching between neighborhoods of nodes being compared. By a set of extensive experiments and case studies, we demonstrate the effectiveness of the proposed model and its instances.

\end{abstract}

\section{Introduction}\label{sec:intro}

How can we measure the similarity between two nodes in a large network?  The de facto established methods are based on the proximity between the two nodes in the network (e.g., shortest distance), or the relative proximity between the two nodes and the other anchor nodes (e.g., SimRank~\cite{jeh2002simrank} and its variations). However, in many applications, such as those where the access to the network is restricted, the proximity or relative proximity methods cannot be used.  Moreover, in different applications we may need different measures, since the  meaning of similarity may vary dramatically.

\begin{example}[Motivation examples]\label{ex:motivation}\em
In Twitter, how can we measure the similarity between two users $u$ and $v$?  In general, we cannot access the whole Twitter network, and thus we cannot compute the shortest distance between $u$ and $v$ or the SimRank~\cite{jeh2002simrank}. Instead, we can run neighborhood queries on selected users as nodes in the network.  Consequently, we can obtain the neighborhoods of $u$ and $v$, which contain the information about the neighbor users (those who follow or being followed by $u$ and $v$, respectively) and the tweets sent by those neighbor users, and, if necessary, the information about the neighbors' neighbors.  Can we measure the similarity between $u$ and $v$ based on the neighborhood information?  This is a fundamental challenge in analyzing many social networks where the network access is constrained.

Even if assuming the access to a whole network, in many applications, similarity may still involve neighborhood information. Consider a social network where nodes are users with their occupations as labels, and edges represent friendship relations, we are interested in finding users whose friends have similar distribution in occupation. Network proximity between two users does not help in measuring their similarity in this context. Instead, we have to measure the similarity based on the neighborhoods of the two nodes being compared.

As another example, in the Yelp network, to compare a leader of a community in Australia and a leader of another community in Vancouver in terms of the structures of the communities they are leading, the proximity between the two leaders or the proximity between the members in the two communities do not help, either.  Instead, we have to compare the communities, which are neighborhoods of the two leaders in the social network, and use effective features, such as the distribution of different types of users and user connection patterns, to design a meaningful similarity measure.

In the above examples and many similar scenarios, proximity measures (e.g., shortest distance) or relative proximity measures (e.g., SimRank or its variations) cannot be applied due to the access restrictions or cannot capture the application meaning of similarity, and thus cannot be used.
\boxend
\end{example}

Several recent studies~\cite{henderson2012rolx,gilpin2013guided} assume a node-feature matrix and measure similarity between nodes accordingly.  However, as illustrated before, different applications may use different node features.  A general framework and a spectrum of similarity measures meeting needs in different applications are still missing.

In this paper, orthogonal to the classical proximity-based and relative proximity-based similarity measures, we tackle the problem of in-network neighborhood-based node similarity measures.  To the best of our knowledge, this problem remains largely untouched.

We make several technical contributions.  First, we propose the notion of neighborhood-based node similarity measure.  This is orthogonal to the well established proximity-based and relative proximity-based node similarity measures.  We give a general parametric model, which captures the critical components in neighborhood-based node similarity measurement.  Different features about neighborhoods can be plugged into our model to produce different instances for various needs in applications.

Second, we examine a series of features popularly used in practice, ranging from maximal common neighborhoods to neighborhood patterns, random walks, and $k$-hops neighbors.  Correspondingly, we derive four instances of the neighborhood-based similarity measures. We explore some desirable and interesting properties of those measures.  We show that except for maximal common neighborhoods based similarity, they can be transformed to metrics.  Moreover, they can all be transformed to normalized similarities.  They all can identify automorphic equivalent nodes.

Third, for the four similarity measures obtained, we analyze the tradeoffs between computational cost and topological matching.  The analysis provides a useful guideline for similarity measure selection in practice.

Last, we conduct extensive experiments to verify the effectiveness of the proposed similarity measures.

The rest of the paper is organized as follows.  Section~\ref{sec:related} reviews the related work. Section~\ref{sec:preliminaries} discusses the preliminaries.  Section~\ref{sec:model} presents out unified parametric model.  Section~\ref{sec:method} derives four instance similarity measures.  Section~\ref{sec:dis} discusses their properties and relations.  Section~\ref{sec:exp} reports an extensive experimental study.  Section~\ref{sec:con} concludes the paper.

\section{Related Work}\label{sec:related}

In-network similarity, which measures similarity between two nodes in a graph, is not a new problem at all, and enjoys many applications.  The dominant methods are  based on either proximity of the two nodes in question (e.g., shortest distance) or the relative proximities between the two nodes and the other anchor nodes.

Node proximity is widely adopted as similarity measures in many data mining and machine learning tasks. Koutra~\textit{et~al.}~\cite{koutra2011unifying} proposed a unified framework of Guilt-by-Association. A wide range of popular proximity measures, such as Personalized PageRank~\cite{jeh2003scaling}, Random Walk with Restart~\cite{tong2006fast} and Generalized Belief Propagation~\cite{yedidia2005constructing}, are special cases or can be approximated by Guilt-by-Association. The basic assumption behind those proximity-based measures is that nodes in a network influence each other through edges and paths. Thus, nodes within the same community are often more similar than those from different communities.  Apparently, proximity-based measures cannot be used to solve the problems demonstrated in Example~\ref{ex:motivation}.

A major group of in-network similarity measures are based on how the nodes in question are connected with the other nodes.   SimRank~\cite{jeh2002simrank} is one of the most popularly used in-network similarity measures. It is based on the principle that two nodes are similar if they are linked to similar nodes. Specifically, to compute the similarity between nodes $u$ and $v$, SimRank aggregates paths between other nodes $w$ and $u$ and between $w$ and $v$, respectively.  SimRank has an undesirable property: if the lengths of paths between $u$ and $v$ are all odd, then $SimRank(u, v) = 0$, that is, $u$ and $v$ cannot be perfectly similar to each other.  To fix this issue, some variations of SimRank were proposed~\cite{jin2011axiomatic,yu2013more}.  Those variations are still based on aggregation of paths between nodes. PathSim~\cite{sun2011pathsim}, a recently proposed similarity measure, employed a similar idea of counting paths following predefined patterns between nodes to derive similarity.  In SimRank and its variations, as well as PathSim, two nodes may not be similar even if their neighborhoods are isomorphic.  Those methods cannot be used to measure the similarity in the applications in Example~\ref{ex:motivation}.

Recently, Henderson~\textit{et~al.}~\cite{henderson2012rolx} proposed RolX, which applied Non-Negative Matrix Factorization to softly cluster nodes in a network. It assumes a node-feature matrix, which is obtained by recursive feature extraction~\cite{henderson2011s}. RolX factorizes the node-feature matrix into a node-role matrix and a role-feature matrix. Then, the node-role matrix can be exploited to compute similarities between nodes. Gilpin~\textit{et~al.}~\cite{gilpin2013guided} introduced supervision to RolX. Since these methods assume a node-feature matrix, they are not specific for in-network neighborhood-based similarity measurement.

In a broader scope, the social science community studied the role of node problem~\cite{lorrain1971structural}. The approaches are mainly based on finding an equivalence relations on nodes so that the nodes can be grouped into equivalent classes. Sparrow~\cite{sparrow1993linear} and Borgatti and Everett~\cite{borgatti1993two} proposed algorithms for finding structural equivalence and regular equivalence classes in which automorphic equivalent nodes are assigned to the same class. However, an equivalence relation on nodes cannot provide node-pair similarities.  Thus, those methods cannot be applied in many data mining and machine learning tasks.

Another line of research related to our study is graph similarity measures. Comparing graphs directly is computationally costly. A major idea of comparing graphs is to convert graphs to feature vectors. For example, Fei and Huan~\cite{fei2008structure} used frequent subgraphs as features of graphs. The most widely studied graph similar measures are graph kernels~\cite{gartner2003graph,kashima2003marginalized,shervashidze2009efficient,shervashidze2011weisfeiler}, which project graphs into a  feature space of finite or infinite dimensionality, and use the inner product of the feature vectors of the two graphs as the similarity score. The features in graph kernels are often substructures of graphs, e.g., walks, subtrees or graphlets.  In Section~\ref{sec:ideas}, we will illustrate why such graph similarity measures cannot be used directly to solve the in-network node similarity problem.

\section{Notations and Preliminaries}\label{sec:preliminaries}


In this paper, we focus on labeled graphs, and use the terms ``graph'' and ``network'', and ``node'' and ``vertex'' interchangeably.

\begin{definition}[Labeled Graph]
A \textbf{labeled graph} (\textbf{graph} for short) is  a tuple $G=\langle V, E, L, \Sigma\rangle $, where $V$ is a set of nodes, $E \subseteq V \times V$ is a set of edges, $\Sigma$ is a set of labels, and $L:V \rightarrow \Sigma$ is a function mapping a node $u \in V$ to a label $l \in \Sigma$.  We often write the set of vertices and the set of edges of a graph $G$ as $V(G)$ and $E(G)$, respectively.
\boxend
\end{definition}

The above definition assumes labels on nodes only. Our methods, however, can be easily extended to graphs with labels on both nodes and edges, or on edges only.


\begin{definition}[Neighborhood]
\label{def:neighborhood}
In a labeled graph $G=\langle V, E, L, \Sigma\rangle$, for a node $u \in V$, the \textbf{$r$-neighborhood} of $u$ $(r \geq 1)$ is a labeled graph $NH_r(u)$=$\langle V_u,E_u,L|_{V_u},\Sigma\rangle $, where $V_u=\{v \mid v \in V, dist(u,v) \leq r\}$, $dist(u,v)$ is the shortest distance from $u$ to $v$, $E_u=\{(v_1,v_2)\mid (v_1,v_2) \in E, v_1, v_2 \in V_u \cup \{u\}\}$, and $L|_{V_u}: V_u \rightarrow \Sigma$ is the function $L$ with restriction on $V_u$. We call $u$ the \textbf{center node} of $NH_r(u)$.  For the sake of brevity, we sometimes omit $r$ and simply write $NH(u)$ when $r$ does not play a role in the discussion.
\boxend
\end{definition}

Some neighborhoods may share some common subgraphs.  To capture common subgraphs shared by neighborhoods of multiple nodes, we define neighborhood patterns.  

\begin{definition}[Neighborhood pattern]\label{def:NP}
A \textbf{neighborhood pattern} is a tuple $\mathcal{G}=\langle H,c\rangle$, where $H$ is a labeled connected graph, and $c \in V(H)$ is called the \textbf{center node} of $\mathcal{G}$.  To keep our presentation concise, we denote by $V(\mathcal{G})=V(H)$ and $E(\mathcal{G})=E(H)$.
\boxend
\end{definition}

Trivially, a neighborhood $\langle NH(u),u\rangle$ itself is a neighborhood pattern. For any subgraph $H$ of $NH(u)$ that contains $u$, that is, $u \in V(H)$, $\langle H, u \rangle$ is a neighborhood pattern that captures part of the neighborhood of $u$.  Please note that Han~\textit{et~al.}~\cite{han2013mining,han2014within} defined a similar concept, pivoted graph.

The matching between a neighborhood pattern and a neighborhood is intuitively defined using subgraph isomorphism.

\begin{definition}[Neighborhood pattern matching]\label{def:nsi}
A neighborhood pattern $\mathcal{G}=\langle H,c\rangle $ is \textbf{neighborhood subgraph isomorphic} (\textbf{NS-isomorphic} for short) to a neighborhood $NH(u)$=$\langle V_u,E_u,L|_{V_u},\Sigma\rangle $, if there exists an injective mapping function $f:V(H) \rightarrow V_u$, such that (1) $f(c)=u$; (2) $\forall v \in V(H)$, $L(v)=L(f(v))$; and (3) $\forall (v_1,v_2) \in E(H)$, $(f(v_1),f(v_2)) \in E_u$.
\end{definition}

\begin{figure}[t]
\centering
\epsfig{file=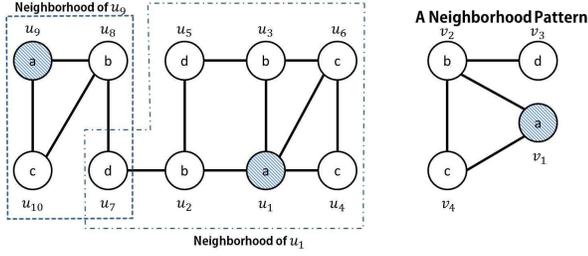, width=80mm}
\caption{Neighborhood subgraph isomorphism.}\label{fig:nsi}
\end{figure}

\begin{example}[NS-isomorphism]\em Figure~\ref{fig:nsi} shows a graph and a neighborhood pattern.  The $2$-neighborhoods of nodes $u_1$ and $u_9$ are highlighted.  The neighborhood pattern is NS-isomorphic to the $2$-neighborhoods of $u_1$ and $u_9$ under the following injective mapping functions, respectively, $f_1(v_1)=u_1$, $f_1(v_2)=u_3$, $f_1(v_3)=u_5$, $f_1(v_4)=u_6$ and $f_2(v_1)=u_9$, $f_2(v_2)=u_8$, $f_2(v_3)=u_7$, $f_2(v_4)=u_{10}$.
\boxend
\end{example}

We define the equivalence of two neighborhood patterns, which is needed when we try to define a set on neighborhood patterns.

\begin{definition}[Neighborhood pattern Equivalence]\label{def:NPE}
Two neighborhood patterns $\mathcal{G}_1=\langle H_1,c_1\rangle $ and $\mathcal{G}_2=\langle H_2,c_2\rangle $ are said to be \textbf{equivalent} to each other if there exists a bijective mapping function $f:V(H_1) \rightarrow V(H_2)$, such that (1) $f(c_1)=c_2$; (2) $\forall v \in V(H_1)$, $L(v)=L(f(v))$; and (3) $\forall v_1, v_2 \in V(H_1), (v_1,v_2) \in E(H_1)$ if and only if $(f(v_1),f(v_2)) \in E(H_2)$.
\boxend
\end{definition}

Automorphic equivalent nodes are often considered as ``identical'' in many applications~\cite{borgatti1993two,lorrain1971structural,jin2011axiomatic}.

\begin{definition}[Automorphic Equivalence]\label{def:AE}
Given a labeled graph $G=\langle V, E, L, \Sigma\rangle$, $\forall u,v \in V$, $u$ and $v$ are \textbf{automorphic equivalent} if there exists a bijective mapping function $f:V \rightarrow V$, such that (1) $f(u)=v \bigwedge f(v)=u$; (2) $\forall v \in V$, $L(v)=L(f(v))$; and (3) $\forall v_1, v_2 \in V, (v_1,v_2) \in E$ if and only if $(f(v_1),f(v_2)) \in E$.
\end{definition}


In addition to neighborhood patterns, another important feature in graphs is walks.

\begin{definition}[Labeled walk]\label{def:labeledWalk}
Given a labeled graph $G= \langle V,E,L,\Sigma \rangle$, a \textbf{walk} $W$ is a sequence of nodes $(u_1,u_2,\ldots,u_m)$, in which $\forall i=1,2,\ldots,m-1, (u_i,u_{i+1}) \in E$. A \textbf{labeled walk} $LW$ is a sequence of labels $(l_1,l_2,\ldots,l_m)$, in which $\forall i=1,2,\ldots,m, l_i \in \Sigma$. The length of $LW$ is $|LW|=m-1$. For a given walk $W=(u_1,u_2,\ldots,u_m)$, the corresponding labeled walk is $L(W)=(L(u_1),L(u_2),\ldots,L(u_m))$, and $W$ is called an \textbf{instance} of $L(W)$.
\boxend
\end{definition}

We also need the concept of product graph, which is widely used in graph kernels.

\begin{definition}[Product Graph]
Given two labeled graphs $G_1$ and $G_2$, the \textbf{product graph} $G_1 \times G_2$ is a labeled graph such that (1) $V({G_1 \times G_2})=\{\langle u',v'\rangle \mid u' \in V(G_1), v' \in V(G_2), L(u')=L(v')\}$; (2) $E({G_1 \times G_2})=\{(\langle u_1,v_1\rangle ,\langle u_2,v_2\rangle) \mid \langle u_1,v_1\rangle, \langle u_2,v_2\rangle  \in V({G_1 \times G_2}), (u_1,u_2) \in E(G_1), (v_1,v_2) \in E(G_2)\}$; and (3) $\forall \langle u',v'\rangle  \in V({G_1 \times G_2}), L(\langle u',v'\rangle )=L(u')=L(v')$.
\boxend
\end{definition}

Please note that a walk may use one node more than once.  An important property of product graph is that every walk (that is, a path) in product graph $G_1 \times G_2$ corresponds to two walks with the same label sequence in $G_1$ and $G_2$, respectively, and vice versa.

Last, we define $k$-hops neighbor set as follows.

\begin{definition}[$k$-Hops Neighbors]\label{def:KN}
For a node $u$ in a labeled graph $G$, the set of \textbf{$k$-hops neighbors} of $u$ is a multi-set of nodes $\{v:w_v \mid w_v \text{ is the number of walks from $u$ to $v$ of length $k$}\}$.
\boxend
\end{definition}

\section{A Unified Parametric Model}
\label{sec:model}

In this section, we first examine why the existing graph comparison methods cannot be used directly to solve the problem. Then, we introduce our unified parametric model. 


\subsection{Neighborhood Comparison versus General Graph Comparison}\label{sec:ideas}

Conceptually, in-network neighborhood-based node similarity measurement is intuitive -- we just need to compare the neighborhoods of two nodes, each neighborhood being a labeled graph.  However, one critical point is that center nodes play special roles in neighborhoods and their comparison.  When we compare two neighborhoods as labeled graphs and try to ``match'' them, the two center nodes have to correspond to each other.

Similar to the general problem of graph comparison, in many application scenarios,  comparing two neighborhoods directly by isomorphism testing is not desirable.  A large network is almost for sure to contain noise and randomness.  The noise edges and vertices severely affect the reliability of measuring similarity by graph matching.  Instead, as a general principle, to measure similarity in a meaningful way, we have to extract structural features from neighborhoods that reflect structural properties of the target nodes. Then, a neighborhood is transformed into a feature vector, and similarity between the induced features can be measured to capture the similarity between the target nodes.

One may wonder whether neighborhood comparison can be tackled using the existing graph comparison methods, such as graph kernels. Unfortunately, general graph comparison methods ignore correspondences between nodes, and thus cannot be used directly in neighborhood comparison.

\begin{figure}[t]
\centering
\epsfig{file=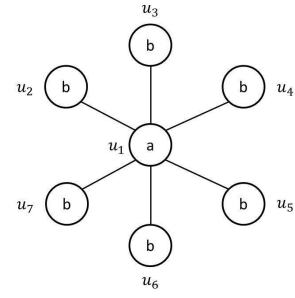, width=40mm}
\caption{An Example}\label{fig:ExpGK}
\end{figure}

\begin{example}[Neighborhood comparison]\em Consider the graph in Figure~\ref{fig:ExpGK}, where $a$ and $b$ are nodes labels. For each node, the $2$-neighborhood is the whole graph. Therefore, graph kernels give the identical similarity score on every pair of nodes.  However, it is easy to see that $u_1$ is uniquely different from the other nodes in the graph.  Moreover, since $u_1$ has a label different from all the others, one may even argue that the similarity between $u_1$ and any other nodes in the graph is very low.  Clearly, this intuition cannot be captured by applying a graph comparison method directly on the neighborhoods.
\boxend
\end{example}

Although neighborhood comparison is very different from general graph comparison, many well established ideas in general graph comparison are very useful in neighborhood comparison.  In the rest of the paper, we will use structural features involving center nodes to assess in-network node similarity.  We will show that the desirable structural features can be obtained by putting more constraints on features extracted for general graph comparison.  We will answer some interesting questions related to general graph comparison, such as whether we can have kernels on nodes within a network analogous to kernels on graph and how those constraints affect the computational cost of similarity computation.

\subsection{A Unified Parametric Framework}

How can we measure the similarity between two target nodes based on their neighborhoods?  We assume that similarity is based on the structural features of the target nodes, that is, the topological structures of the neighborhoods and how labels are distributed in the neighborhoods.

There are many possible structural features that can be used in in-network node similarity measures. Importantly, different applications may use different structural features and assemble features in different ways.  Instead of proposing many yet another similarity measures, we are interested in a fundamental scientific problem: is there a general framework about the mechanism measuring neighborhood-based similarity? The general framework should be parametric -- specific features and specific methods of feature assembling are configurable parameters.  Moreover, the framework should generalize a series of meaningful similarity measures.

In general, we make the following basic assumptions about measuring in-network neighborhood-based node similarity.

\begin{assumption}[Features] In-network neighborhood-based node similarity can be measured by aggregation over a set of features.

{\noindent\sc Rationale. }\em We assume that there exists a set of features $\mathcal{F}$ such that the overall similarity between two nodes is an aggregation of the similarity scores between the two nodes over individual features $\Gamma \in \mathcal{F}$. This is analogous to Minkowski distance where the overall distance is an aggregation of the distances in all individual dimensions (i.e., $\big(\sum_{\text{dimension } d} (\text{distance on }d)^p\big)^{1/p}$).  This is a natural and popularly used way to measure similarity and distance in a multidimensional space in a divide and conquer manner.
\boxend
\end{assumption}

Based on the above assumption, we propose the following general parametric model.

\begin{definition}[A general similarity model]\label{defn:general-model}
Given a graph $G$ and two nodes $u, v \in V(G)$, the similarity between $u$ and $v$ is measured by
\begin{equation}\label{eq:general}
    Sim(u,v)=\mathop{\Huge{\Phi}}\limits_{\Gamma \in \mathcal{F}}\Big(\mathop{\Huge{\Xi}}\limits_{\small\begin{array}{c}
    s \text{ is an element of } NH(u) \\
    t \text{ is an element of } NH(v)
        \end{array} } {\sigma(u,v,s,t,\Gamma)}\Big)
\end{equation}
where four parameters are used
\begin{itemize}
\item $\mathcal{F}$ is a set of features used in the similarity measure;
\item $\sigma$ is a scoring function returning the similarity score between $u$ and $v$ in the evidence of elements $s$ and $t$ on feature $\Gamma \in \mathcal{F}$, where variables $s$ and $t$ are elements, that is, either nodes or edges, in $NH(u)$ and $NH(v)$, respectively;
\item $\Xi$ is an aggregate function defining how the scores with respect to various element pairs are summarized, and
\item $\Phi$ is another aggregate function specifying how the features are assembled in measuring the overall similarity.
\boxend
\end{itemize}
\end{definition}

In Equation~\ref{eq:general}, we aggregate the similarity scores over all features $\Gamma \in \mathcal{F}$ to derive the overall similarity between two nodes $u$ and $v$.  To derive the similarity score on each feature $\Gamma$, we consider every pair of elements in the neighborhoods of $u$ and $v$, where the elements used are either nodes or edges, and may vary from one measure to another.  For elements $s$ from $NH(u)$ and $t$ from $NH(v)$, we calculate a similarity score $\sigma(u, v, s, t, \Gamma)$ that is the similarity between $u$ and $v$ reflected by $s$ and $t$ on feature $\Gamma$.  The similarity score between $u$ and $v$ on feature $\Gamma$ is an aggregation of the similarity scores over all pairs $(s, t)$.  In the next section, we will present four instances of the model, which will illustrate the possible implementations and flexibility of the model clearly.


\section{Four Instance Similarity Measures}
\label{sec:method}

In this section, we present four instances of the general model (Definition~\ref{defn:general-model}) as interesting and practically useful similarity measures.  Unlike~\cite{henderson2011s,henderson2012rolx,gilpin2013guided}, where features are basically statistics of nodes' neighbourhoods, we focus on graph substructure features like subgraphs and walks. The reason we choose this kind of features is that they have better interpretability. By checking features of $u$ and $v$, we know how many common substructures, such as subgraphs, subtrees, and walks, are shared by their neighborhoods.  This information provides an intuitive explanation why $u$ and $v$ are similar. While for statistical features in~\cite{henderson2011s, henderson2012rolx,gilpin2013guided}, it is not easy to find what they really capture in the physical world.

\subsection{Similarity by Maximum Common Neighborhood Pattern (MCNP)}\label{ssec:SimMCNP}

A simple idea to evaluate the similarity between two nodes is to measure the size of the largest common subgraph shared by the neighborhoods of the two nodes.

\begin{definition}[MCNP similarity]
Given a network $G$, for nodes $u, v \in V(G)$, a neighborhood pattern $\mathcal{G}=\langle H, c \rangle$ is a \textbf{maximal common neighborhood pattern} (\textbf{MCNP} for short) of $u$ and $v$ if $\mathcal{G}$ is NS-isomorphic to both $NH(u)$ and $NH(v)$, and $|E(\mathcal{G})|$ is maximized.

The \textbf{MCNP similarity} between $u$ and $v$ is $$SimMCNP(u, v) = |E(\mathcal{G})|,$$ where $\mathcal{G}$ is a MCNP of $u$ and $v$.
\boxend
\end{definition}

MCNP similarity is meaningful in many applications.  For example, in a friendship network $G$, the similarity between two nodes $u$ and $v$ can be measured by the size of a maximal common subgraph shared by the neighborhoods of $u$ and $v$, which has the largest number of friendship links.

\begin{example}[MCNP similarity]\em
In Figure~\ref{fig:nsi}, it happens that the neighborhood pattern is a MCNP of $u_1$ and $u_9$.  The MCNP similarity between $u_1$ and $u_9$ is $4$.
\boxend
\end{example}

MCNP similarity is an instance of the general similarity model in Definition~\ref{defn:general-model}.  Specifically, given a graph $G$ and a node $u$, let
\begin{equation}
\begin{array}{rcl}
SNP(u) & = &
\{\mathcal{G}=\langle H, c\rangle \mid \mathcal{G} \text{ is NS-isomorphic to } NH(u) \\
& &\wedge\; H \text{ is connected}\}\\
\end{array}\label{eq:SNP}
\end{equation}
be the set of neighborhood patterns that are NS-isomorphic to the neighborhood of $u$.  For two nodes $u, v \in V(G)$, a neighborhood pattern $\mathcal{G}=\langle H, c \rangle \in SNP(u) \cap SNP(v)$ if there exist $\mathcal{G}_1= \langle H_1, c_1\rangle \in SNP(u)$ and $\mathcal{G}_2= \langle H_2, c_2\rangle \in SNP(v)$ such that $H$ and $H_1$ are isomorphic and $c$ and $c_1$ correspond to each other in the isomorphism bijective mapping, $H$ and $H_2$ are isomorphic and $c$ and $c_2$ correspond to each other in the isomorphism bijective mapping.  Apparently, we have the following.

\begin{definition}[Pseudo-Inverse]
Let $f:A \rightarrow B$ be an injective function. A function $f^{-1}:B \rightarrow (A \bigcup \{nil\})$ is a \textbf{pseudo-inverse} of $f$, if for each $b \in B$,
$$
    f^{-1}(b)=
    \begin{cases}
        nil & \forall x \in A, f(x) \neq b \\
        a & a \in A \bigwedge f(a)=b
    \end{cases}\nonumber
$$
\end{definition}

\begin{theorem}\label{thm:MCNP}
Given a graph $G$, for two nodes $u, v \in V(G)$,
$$
\begin{array}{rl}SimMCNP(u, v)=&
\max_{
\begin{subarray}{l}
\mathcal{G}=\langle H, c\rangle \in SNP(u) \cap SNP(v)\\
\text{w.r.t.\ injective mapping}\\
\text{functions } f_u: V(H) \rightarrow V(NH(u))\\
\text{and }f_v: V(H) \rightarrow V(NH(v))
\end{subarray}}\\
&  \Big(
\sum_{
\begin{subarray}{c}
(x, y) \in E(NH(u)) \\ (x', y') \in E(NH(v))
\end{subarray}}\\
& I\big(f_u^{-1}(x)=f_v^{-1}(x')\neq nil \wedge \\
& f_v^{-1}(y)=f_v^{-1}(y')\neq nil\big)\Big),
\end{array}$$
where $I(P)$ is an indicator function, which takes $1$ if $P$ is true and $0$ otherwise, and $f_u^{-1}$ and $f_v^{-1}$ are pseudo-inverses of $f_u$ and $f_v$, respectively.
\boxend
\end{theorem}

In Theorem~\ref{thm:MCNP}, $f_u(c)=u$ and $f_v(c)=v$ because $\mathcal{G}$ is NS-isomorphic to both $NH(u)$ and $NH(v)$. This condition actually is the key property that general graph comparison methods do not guarantee. It ensures that when we assess the similarity between $NH(u)$ and $NH(v)$, $u$ and $v$ always correspond to each other.

\subsection{Similarity by Neighborhood Patterns}\label{ssec:SimNP}

In some application scenarios~\cite{han2013mining}, one may want to use patterns as features, and measure the similarity of two nodes by comparing the common features occurring in their neighborhoods. This is analogous to, for example, using frequent subgraphs as features in graph comparison~\cite{fei2008structure}.


\begin{definition}[NP similarity]\label{def:SimNP}
Given a network $G$ and a set of a neighborhood patterns $S$, for nodes $u, v \in V(G)$, the \textbf{neighborhood pattern similarity} (\textbf{NP similarity} for short) between $u$ and $v$ is $$SimNP(u, v) = \sum_{\mathcal{G} \in \big(S \cap SNP(u) \cap SNP(v)\big)}{|E(\mathcal{G})|}$$ where $SNP()$ is defined in Equation~\ref{eq:SNP}.
\boxend
\end{definition}

NP similarity enables users with the full flexibility to specify neighborhood patterns as features.  There are many ways to select $S$.  For example, we may use neighborhood patterns that are frequent in neighborhoods of all nodes in the whole graph~\cite{han2013mining}. We may set $S$ as a set of predefined subgraphs, such as Meta-Path~\cite{sun2011pathsim}. It is also feasible to use graphlets~\cite{shervashidze2009efficient} to form $S$.

The NP similarity is an inner product of two feature vectors, that is, $SimNP(u,v)=vec_{NP}(u)^Tvec_{NP}(v)$, where the $j$-th element of the feature vector of node $u$ is
\begin{equation}
    vec_{NP}(u)_j =
    \begin{cases}
        \sqrt{|E({\mathcal{G}_j})|} & \textit{$\mathcal{G}_j$ is NS-isomorphic to NH(u)} \\
        0 & otherwise
    \end{cases}\nonumber
\end{equation}
and $\mathcal{G}_j$ is the $j$-th neighborhood pattern in $S$.

It is easy to show that $SimNP$ is an instance of the general similarity model in Definition~\ref{defn:general-model}.

\begin{theorem}
Given a graph $G$ and a set of neighborhood patterns $S$, for two nodes $u, v \in V(G)$,
$$
\begin{array}{rl}SimMCNP(u, v)=&
\sum_{
\begin{subarray}{l}
\mathcal{G}=\langle H, c\rangle \in SNP(u) \cap SNP(v) \cap S\\
\text{w.r.t.\ injective mapping}\\
\text{functions } f_u: V(H) \rightarrow V(NH(u))\\
\text{and }f_v: V(H) \rightarrow V(NH(v))
\end{subarray}}\\
&  \Big(
\sum_{
\begin{subarray}{c}
(x, y) \in E(NH(u)) \\ (x', y') \in E(NH(v))
\end{subarray}}\\
& I\big(f_u^{-1}(x)=f_v^{-1}(x')\neq nil \wedge \\
& f_v^{-1}(y)=f_v^{-1}(y')\neq nil \big)\Big)
\end{array}$$
where $f_u^{-1}$ and $f_v^{-1}$ are pseudo-inverses of $f_u$ and $f_v$, respectively.
\boxend
\end{theorem}

Similar to $SimMCNP$, when using $SimNP$ to compute the similarity of $NH(u)$ and $NH(v)$, $u$ and $v$ always correspond to each other.

\subsection{Similarity by Labeled Walks}\label{ssec:SimLW}

Walks are popularly used as structural features in network analysis.  Can we measure the similarity between two nodes based on the walks in their neighborhoods?


\begin{definition}[LW feature vector]\label{def:LWFV}
Given a graph $G= \langle V,E,L,\Sigma \rangle$ and a parameter $r$, the \textbf{labeled walk feature set} is $$S=\{LW=(l_1,l_2,\ldots,l_m) \mid 1 \leq |LW| \leq r, \forall i=1,2,\ldots,m, l_i \in \Sigma\}.$$ The \textbf{labeled walk feature vector} (\textbf{LW feature vector} for short) of node $u\in V$ is a $|S|$-dimensional vector $vec_{LW}(u)$, in which $$vec_{LW}(u)_j=\sqrt{\lambda_{|LW_j|}}*|\{W=(u_1,u_2,\ldots,u_m) \mid u_1=u, L(W)=LW_j\}|$$ $LW_j \in S$, and $\lambda_1, \lambda_2, \ldots, \lambda_r$ are parameters such that $\forall i=1,2,\ldots,r$, $\lambda_i \geq 0$ and $\sum_{i=1}^{r}{\lambda_i}=1$.
\boxend
\end{definition}

A user can use parameters $\lambda_i$'s to control how walks of different lengths contribute to the similarity.  In many applications, shorter walks are regarded contributing more to the final similarity score than longer ones~\cite{gartner2003graph}.

Apparently, the set of all possible labeled walks of length up to $r$ is finite provided that the set of labels is finite. Using LW feature vectors, we define labeled walk similarity.

\begin{definition}[LW similarity]\label{def:SimLW}
Given a graph $G= \langle V,E,L,\Sigma \rangle$, for nodes $u,v \in V(G)$, the \textbf{labeled walk similarity} (\textbf{LW similarity} for short) between $u$ and $v$ is $$SimLW(u,v)=vec_{LW}(u)^Tvec_{LW}(v)=\sum_{LW_j \in S}{vec_{LW}(u)_j*vec_{LW}(v)_j}.$$
\boxend
\end{definition}

We can show that $SimLW$ is also an instance of our general similarity model in Definition~\ref{defn:general-model}.

\begin{theorem}\label{th:LW} Denote by $A_{G \times G}$ the adjacency matrix of the product graph $G \times G$.
    $$SimLW(u,v)=\sum_{i=1}^{r}\mathop{\sum}\limits_{\begin{subarray}{l}
    u' \in V(NH(u)) \\
    v' \in V(NH(v))
        \end{subarray}}{\lambda_i A_{G \times G}^{i}(\langle u,v\rangle ,\langle u',v'\rangle )}$$
\proof\em
Denote by $W_{u \rightarrow}$ a walk starting at node $u$, and by $W_{u \rightarrow v}$ a walk from $u$ to $v$.

\begin{small}
\begin{equation*}
\begin{split}
&\indent SimLW(u,v) \\
&= \mathop{\sum}\limits_{LW_j \in S}{vec_{LW}(u)_{j}*vec_{LW}(v)_{j}} \\
&= \mathop{\sum}\limits_{i=1}^{r}\mathop{\sum}\limits_{j:|LW_j|=i}{vec_{LW}(u)_{j}*vec_{LW}(v)_{j}} \\
&= \mathop{\sum}\limits_{i=1}^{r} \mathop{\sum}\limits_{j:|LW_j|=i}\lambda_i{|\{W_{u \rightarrow} \mid L(W_{u \rightarrow})=LW_j\}|*|\{W_{v \rightarrow} \mid L(W_{v \rightarrow})=LW_j\}|} \\
&= \mathop{\sum}\limits_{i=1}^{r} \mathop{\sum}\limits_{j:|LW_j|=i}\lambda_i\mathop{\sum}\limits_{\substack{u' \in \mathrlap{V(NH(u))} \\ v' \in \mathrlap{V(NH(v))}}}{|\{W_{u \rightarrow u'} \mid L(W_{u \rightarrow u'})=LW_j\}|*|\{W_{v \rightarrow v'} \mid L(W_{v \rightarrow v'})=LW_j\}|} \\
&= \mathop{\sum}\limits_{i=1}^{r}\mathop{\sum}\limits_{\begin{subarray}{l}
        u' \in V(NH(u)) \\
        v' \in V(NH(v))
    \end{subarray}}{\lambda_i \mathop{\sum}\limits_{j:|LW_j|=i}{|\{W_{u \rightarrow u'} \mid L(W_{u \rightarrow u'})=LW_j\}|*|\{W_{v \rightarrow v'} \mid L(W_{v \rightarrow v'})=LW_j\}|}} \\
&= \mathop{\sum}\limits_{i=1}^{r}\mathop{\sum}\limits_{\begin{subarray}{l}
        u' \in V(NH(u)) \\
        v' \in V(NH(v))
    \end{subarray}}{\lambda_i \mathop{\sum}\limits_{j:|LW_j|=i}{|\{W_{\langle u,v\rangle  \rightarrow \langle u',v'\rangle } \mid L(W_{\langle u,v\rangle  \rightarrow \langle u',v'\rangle })=LW_j\}|}} \\
&= \mathop{\sum}\limits_{i=1}^{r}\mathop{\sum}\limits_{\begin{subarray}{l}
        u' \in V(NH(u)) \\
        v' \in V(NH(v))
    \end{subarray}}{\lambda_i |\{W_{\langle u,v\rangle  \rightarrow \langle u',v'\rangle } \mid |W_{\langle u,v\rangle  \rightarrow \langle u',v'\rangle }|=i\}|} \\
&= \mathop{\sum}\limits_{i=1}^{r}\mathop{\sum}\limits_{\begin{subarray}{l}
                    u' \in V(NH(u)) \\
                    v' \in V(NH(v))
                \end{subarray}}{\lambda_i A_{G \times G}^{i}(\langle u,v\rangle ,\langle u',v'\rangle )}
\end{split}
\end{equation*}
\end{small}

Here $|\{W_{u \rightarrow u'} \mid L(W)=LW_j\}|*|\{W_{v \rightarrow v'} \mid L(W)=LW_j\}|=|\{W_{\langle u,v\rangle  \rightarrow \langle u',v'\rangle } \mid L(W)=LW_j\}|$ is due to the property of product graph.
\boxend
\end{theorem}

As shown in Theorem~\ref{th:LW}, the features are labeled walks of length $i$, the elements for similarity computing are nodes in neighborhoods, and the scoring function is $\lambda_i A_{G \times G}^{i}(\langle u,v\rangle ,\langle u',v'\rangle)$.

$SimLW$ is related to walk based graph kernels: the features used are labeled walks.  However, in $SimLW$, we have more constraints on instances of labeled walks (i.e., walks) used. First, all walks selected in a neighborhood for computing similarity must have the center node as the start point.  This constraint not only ensures that all features are related to the center nodes, but also guarantees that the center nodes always play the same role in features. Second, we only use walks of limited lengths. This constraint ensures that all walks used for computing similarity are within the neighborhoods of the center nodes.

\subsection{Similarity by $k$-hops Neighbors}\label{ssec:SimKN}

$SimLW$ uses labeled walks as features. The number of possible label walks may be huge, in $O(|\Sigma|^r)$, where $\Sigma$ is the set of labels. To reduce the number of features, we can relax the labeled walk features by only considering the endpoints of walks.

\begin{definition}[$k$-hops Neighbor Feature Vector]\label{def:KNFV}
Given a graph $G= \langle V,E,L,\Sigma \rangle$ and a parameter $r$, the \textbf{$k$-hops neighbor feature vector} of node $u$ is an $r|\Sigma|$-dimensional vector $vec_{KN}(u)=(\sqrt{\lambda_1}{n_u^1}^T,\sqrt{\lambda_2}{n_u^2}^T,\ldots,\sqrt{\lambda_r}{n_u^r}^T)$, where $\forall i=1,2,\ldots,r$, $n_u^i$ is a $|\Sigma|$-dimensional vector that $(n_u^i)_l$ is the sum of multiplicities of nodes with label $l$ in $u$'s $i$-hops neighbor set, and $\lambda_1, \lambda_2, \ldots, \lambda_r$ are parameters such that $\forall i=1,2,\ldots,r$, $\lambda_i \geq 0$ and $\sum_{i=1}^{r}{\lambda_i}=1$.
\boxend
\end{definition}

The multiplicity of a node $v$ in $u$'s $k$-hops neighbor set is the number of walks from $u$ to $v$ of length $k$, according to Definition~\ref{def:KN}. Thus, we obtain  $$n_u^k=\sum_{v \in Neighbor(u)}{n_v^{k-1}}$$ Therefore, $u$'s $k$-hops neighbor features can also be explained as recursive features~\cite{henderson2011s} that are recursively built by aggregations of $u$'s neighbors' information.

Based on the $k$-hops neighbor feature vectors, we can define the $k$-hops neighbor similarity.
\begin{definition}[KN Similarity]\label{def:SimKN}
Given a graph $G= \langle V,E,L,\Sigma \rangle$, for nodes $u,v \in V(G)$, the \textbf{$k$-hops neighbor similarity} (\textbf{KN similarity} for short) between $u$ and $v$ is $$SimKN(u,v)=vec_{KN}(u)^Tvec_{KN}(v)$$
\boxend
\end{definition}

The roles of parameters $\lambda_i$'s are the same as in $SimLW$. In a way similar to the proof of Theorem~\ref{th:LW}, we can also show that $SimKN$ is an instance of our general similarity model in Definition~\ref{defn:general-model}.

\begin{theorem}\label{th:KN} Denote by $A_G$ the adjacency matrix of $G$.
    $$\begin{array}{l}
    SimKN(u,v) = \\
    \sum_{i=1}^{r}\mathop{\sum}\limits_{\begin{subarray}{l}
                      u' \in V(NH(u)) \\
                      v' \in V(NH(v))
                      \end{subarray}}\lambda_i A_G^i{(u,u')}A_G^i{(v,v')}*I(L(u)=L(v) \wedge  L(u')=L(v'))
     \end{array}$$
\boxend
\end{theorem}

Similar to $SimLW$, our constraints on features make sure that the center nodes are always matched to each other when computing similarity.

\section{Relations among the Four Similarities}
\label{sec:dis}
In this section we discuss important properties of the four similarities proposed in Section~\ref{sec:method}, as well as the relations among them.

\subsection{Normalization}

In many applications, such as K-means clustering, where a metric is needed, it is highly desirable that a similarity measure can be normalized to a distance metric. Except for $SimMCNP$, the similarities proposed can be normalized to metrics.

\begin{theorem}[Metric]
Let $Sim(\cdot,\cdot)$ be $SimNP(\cdot,\cdot)$, $SimLW(\cdot,\cdot)$ or $SimKN(\cdot,\cdot)$.  Define
\begin{equation}\label{eq:metric}
        Dist(u,v) = \sqrt{Sim(u,u)-2Sim(u,v)+Sim(v,v)}
\end{equation}
$Dist(\cdot, \cdot)$ is a metric.

\proof\em
According to the definitions, $SimNP(\cdot,\cdot)$, $SimLW(\cdot,\cdot)$ and $SimKN(\cdot,\cdot)$ are all inner products in their corresponding feature spaces. Immediately, we can plug each of those inner products into Equation~\ref{eq:metric} to obtain an Euclidean distance between nodes~\cite{deza2009encyclopedia}, which is a metric.
\boxend
\end{theorem}

Moreover, in many applications, it is desirable to have a similarity that can recognize automorphic equivalence~\cite{borgatti1993two,lorrain1971structural,jin2011axiomatic}.  However, some well-known similarity measures, such as SimRank~\cite{jeh2002simrank}, do not have this properly.  Nicely, all the four measures proposed here can be easily normalized to have the property.


\begin{theorem}[Normalized similarity] Let $Sim(\cdot,\cdot)$ be $SimMCNP(\cdot,\cdot)$, $SimNP(\cdot,\cdot)$, $SimLW(\cdot,\cdot)$ or $SimKN(\cdot,\cdot)$. Define
\begin{equation}\label{eq:NSim}
       NSim(u,v)=\frac{Sim(u,v)}{Sim(u,u)+Sim(v,v)-Sim(u,v)}
\end{equation}
Then, $0 \leq NSim(u, v) \leq 1$ and if $u$ and $v$ are automorphic equivalent, $NSim(u, v)=1$.

\proof\em
Obviously, $\forall u,v \in V$, $Sim(u,u) \geq Sim(u,v) \geq 0$. Thus, $Sim(u,u)+Sim(v,v)-Sim(u,v) \geq Sim(u,u) \geq Sim(u,v)$. Then, we have, $\forall u,v \in V$, $0 \leq NSim(u,v) \leq 1$.

If $u$ and $v$ are automorphic equivalent, $Sim(u,u)=Sim(u,v)=Sim(v,v)$ because $NH(u)$ and $NH(v)$ must be isomorphic. Thus, $NSim(u,v)=\frac{Sim(u,u)}{Sim(u,u)+Sim(u,u)-Sim(u,u)}=1$.
\boxend
\end{theorem}

We can further obtain a generalized Jaccard similarity measure~\cite{Gregson1975} by plugging $SimNP(\cdot,\cdot)$ into Equation~\ref{eq:NSim}.

\begin{theorem}[Generalized Jaccard similarity]\label{thm:jaccard} $NSim(\cdot,\cdot)=\frac{SimNP(\cdot,\cdot)}{SimNP(\cdot,\cdot)+SimNP(\cdot,\cdot)-SimNP(\cdot,\cdot)}$
is a generalized Jaccard similarity measure.

\proof\em
The generalized Jaccard similarity~\cite{Gregson1975} has the form $J(x,y)=\frac{\sum_{i}{\min{(x_i,y_i)}}}{\sum_{i}{\max{(x_i,y_i)}}}$, where $x=(x_1,x_2,...,x_n)$ and $y=(y_1,y_2,...,y_n)$ are two vectors and $x_i,y_i \geq 0$ are real numbers.  For each node $u \in V$, we create a vector $x(u)$ with the same dimensionality as $vec_{NP}(u)$ and set $x(u)_i=vec_{NP}(u)_i^2$. Then, for two nodes $u$ and $v$, straightforwardly, $\min{(x(u)_i,x(v)_i)}=vec_{NP}(u)_ivec_{NP}(v)_i$ and $\max{(x(u)_i,x(v)_i)}=vec_{NP}(u)_i^2+vec_{NP}(v)_i^2-vec_{NP}(u)_ivec_{NP}(v)_i$. Therefore, $J(x(u),x(v))=\frac{SimNP(u,v)}{SimNP(u,u)+SimNP(v,v)-SimNP(u,v)}=NSim(u,v)$.
\boxend
\end{theorem}


\subsection{Relations between $SimLW$ and $SimKN$}\label{ssec:relation}

Both $SimLW$ and $SimKN$ use walks as the structural features.  The difference is that $SimKN$ only considers the end points while $SimLW$ considers the whole labeled walks.  $SimKN$ considers a subset of information that is considered by $SimLW$.  Intuitively, if two nodes are similar in $SimLW$, they must be also similar in $SimKN$, but not the other way.  We establish this relationship formally.

\begin{theorem}[$SimLw$ and $SimKN$]\label{th:upper}
Under the same parameters $r$ and $\lambda_i$'s, $SimLW(u,v) \leq SimKN(u,v)$.
\proof\em
We only need to prove $A_{G \times G}^{i}(\langle u,v\rangle ,\langle u',v'\rangle ) \leq A_G^i{(u,u')}A_G^i{(v,v')}*I\{L(u)=L(v) \bigwedge  L(u')=L(v')\}$.  Two cases may arise.

Case 1: $L(u)=L(v)$ and $L(u')=L(v')$. According to the property of product graphs, $A_{G \times G}^{i}(\langle u,v\rangle ,\langle u',v'\rangle )$ is the number of walk pairs $(W_1, W_2)$ in $G$ such that $W_1$ is from $u$ to $u'$ and has length $i$, $W_2$ is from $v$ to $v'$ and has length $i$, and the label sequences of the two walks are the same. Clearly, $A_{G \times G}^{i}(\langle u,v\rangle ,\langle u',v'\rangle ) \leq A_G^i{(u,u')}A_G^i{(v,v')}$ holds because $A_G^i{(u,u')}$ is the total number of walks of length $i$ from $u$ to $u'$.

Case 2: $L(u)\neq L(v)$ or $L(u') \neq L(v')$. In this case, $A_{G \times G}^{i}(\langle u,v\rangle ,\langle u',v'\rangle )=0$. Moreover, because both $A_G^i{(u,u')}$ and $A_G^i{(v,v')}$ are non-negative, $A_{G \times G}^{i}(\langle u,v\rangle ,\langle u',v'\rangle ) \leq A_G^i{(u,u')}A_G^i{(v,v')}$ holds immediately.
\boxend
\end{theorem}

\subsection{Computational Cost}

We first show that both $SimMCNP$ and $SimNP$ intractable.  Then, we discuss the computation cost of $SimLW$ and $SimKN$ using preprocessing, and thus show that $SimLW$ and $SimKN$ are tractable.  Last, we discuss the tradeoff among the four similarity measures about features used and computational cost.

\subsubsection{Intractability of SimMCNP and SimNP}


\begin{theorem}[$SimMCNP$]\label{th:MCNG}
Given a labeled graph $G$, nodes $u, v \in V(G)$, $r\geq 1$, and $l >0$, the decision problem on whether $SimMCNP(u,v) \geq l$ is NP-hard.

\proof\em
We prove by a reduction from the NP-hard clique problem, which is to decide if a given clique $G_2$ is subgraph isomorphic to a given graph $G_1$.

Suppose we have an oracle that, given a labeled graph $G=\langle V,E,L,\Sigma\rangle$, two nodes $u \in V$ and $v \in V$, and $l>0$, when the constant $r \geq 1$, can return if $SimMCNP(u,v) \geq l$ in polynomial time with respect to the size of $G$.

For any instance $(G_1=\langle V_1,E_1\rangle ,G_2=\langle V_2,E_2\rangle )$ of the clique problem, where $G_2$ is a clique graph having $|V_2|$ nodes, first we assign one same label to all nodes. Then, we create a new labeled graph $G=G_1 \bigcup G_2$. Note that the nodes in $V_2$ are automorphic equivalent. Thus, for any two nodes $u_1 \in V_2$ and $u_2 \in V_2$, and an arbitrary other node $v$, $SimMCNP(u_1,v)=SimMCNP(u_2,v)$. So we pick an arbitrary node $u \in V_2$. Then, for each node $v \in V_1$, we call the oracle to see if $SimMCNP(u,v) \geq |E_2|$. Clearly $G_2$ is subgraph isomorphic to $G_1$ if and only if there exists at least one node $v \in V_1$ such that $SimMCNP(u,v) \geq |E_2|$. So we can call the oracle at most $|V_1|$ times to decide if $G_1$ has a clique with size $|V_2|$. Each calling of the oracle takes polynomial time with respect to the size of $G$, which means in total we can use polynomial time to decide the clique probelm. This means deciding if $SimMCNP(u,v) \geq l$ is NP-hard.
\boxend
\end{theorem}

\begin{theorem}[$SimNP$]\label{th:NP}
Given a graph $G$, a neighborhood pattern set $S$, nodes $u, v \in V(G)$, $r\geq 1$, and $l >0$, the decision problem on whether $SimNP(u,v) \geq l$ is NP-hard.

\proof\em
The proof is similar to the proof of Theorem~\ref{th:MCNG}. For an instance $(G_1=\langle V_1,E_1\rangle ,G_2=\langle V_2,E_2\rangle )$ of the clique problem, we just create an instance of computing $SimNP$ by setting the neighborhood pattern set $S=\{G_2\}$. 
\boxend
\end{theorem}


\subsubsection{Preprocessing for $SimLW$ and $SimKN$}

Both $SimLW$ and $SimKN$ are inner products of feature vectors.  Therefore, to speed up the computation of those similarities, we can materialize feature vectors for all nodes offline.  Please note that $r$, the radius of neighborhood, is a predefined parameter. Due to the well recognized small world phenomenon~\cite{kleinberg2000small}, the parameter $r$ should be set to a small number like $2$. Otherwise, a neighborhood can easily involve a large portion of the whole graph. Normally $|\Sigma| \ll |V|$.  Let us investigate the cost of processing for $SimLW$ and $SimKN$.




\paragraph{Preprocessing for $SimLW$}

We do not need to explicitly build the adjacent matrix of $G \times G$ and compute
$A_{G\times G}^i$ $(1 \leq i \leq r)$, which takes $O(|V|^6)$ time.  We can first index all possible labeled walks considered, and search the neighborhood of each node to count the frequencies of the labeled walks. A straightforward implementation is to conduct a breadth-first search on the neighborhood of each node to enumerate all labeled walks.

The number of walks of length $i$, starting from a center node $u$ and ending at a node $v \in V_u$, is no more than $|V_u|^{i-1}$.
The cost of computing the labeled walk feature vector $vec_{LW}(u)$ is $O(\sum_{i=1}^{r-1}{|V_u|^{i-1}|E_u|})=O(|V_u|^{r-2}|E_u|)$. Therefore, the total cost of computing all labeled walk feature vectors is $O(\sum_{u \in V}{|V_u|^{r-2}|E_u|})$, which is definitely bounded by $O(|V|^{r-1}|E|)$.

\paragraph{Preprocessing for $SimKN$}
To analyze the time complexity of materializing $k$-hops neighbor features, we follow the recursive computation of $vec_{KN}(u)$ introduced in section~\ref{ssec:SimKN}. Clearly, computing all $n_u^1$'s takes $O(E)$ time. Computing all $n_u^{i}=\sum_{v \in N(u)}{n_v^{i-1}}$ for $i \ge 1$ takes $O(|\Sigma||E|)$ time. In total the overall time complexity of preprocessing is only $O(r|\Sigma||E|)$ time.

\bigskip
Please note that the above methods using preprocessing are just to show the tractability of $SimLW$ and $SimKN$.  Efficient and scalable algorithms computing the similarity measures are beyond the capacity of this paper, and are reserved as future work.

\subsubsection{Tradeoffs}\label{subsec:tradeoff}

$SimMCNP$ is a very strict similarity measure -- two nodes are similar if and only if their neighborhoods share a large common subgraph.  $SimNP$ poses a weaker requirement.  Two nodes are similar if some selected patterns appear in the neighborhoods of the nodes.  The neighborhoods of the nodes may have very different topological structures and do not have to share a large common subgraph.  $SimLW$ is even weaker since it only considers labeled walks.  Two neighborhoods with very different topological structures may still share many common labeled walks.  As indicated by Theorem~\ref{th:upper}, $SimKN$ is even weaker than $SimLW$.  Roughly speaking, the strictness of requiring topological matching between neigbhorhoods in the similarity measures decreases from $SimMCNP$ to $SimNP$, to $SimLW$ and to $SimKN$.

At the same time, computing $SimMCNP$ and $SimNP$ are much harder than computing $SimLW$ and $SimKN$. Moreover, $SimLW$ is more costly than $SimKN$.  Therefore, the four similarity measures present meaningful tradeoffs between strictness of matching and computational cost.

\section{Empirical Results}\label{sec:exp}

In this section, we report a series of empirical studies on the four similarities proposed using both synthetic data and real data. First, we use a synthetic data set to illustrate some interesting pairs of nodes whose neighborhoods are similar in one way or another.  Second, we employ three data analytic tasks, namely similar authors search, anomaly nodes ranking, and nodes clustering, to illustrate the effectness of our proposed similarity measures in real applications.

We find that, even the four similarities are all based on structural features from neighborhoods, the semantics of similarity captured by them have some important differences. Therefore, for each similarity, there are some applications suitable. By exploring the experimental results and the semantics captured by each proposed similarity in the tasks, our empirical studies provide a useful guideline on selecting the appropriate similarity measures according to application needs.

\subsection{Experimental Settings}\label{subsec:setting}

Due to the well recognized small world phenomenon~\cite{kleinberg2000small}, the parameter $r$ should be set to a small number.  In our experiments, we set $r=2$.


Because of the high computational cost in $SimMCNP$, we only use it in the experiments reported in Section~\ref{sec:case-study}, where the graph size is the smallest.  For $SimNP$, the set $S$ consists of neighborhood patterns $\mathcal{G}$ such that (1) $\mathcal{G}$ is connected; (2) $\mathcal{G}$ has at most $n$ edges; and (3) $\mathcal{G}$ is NS-isomorphic to the neighborhoods of at least $\tau$ nodes in $G$. The values of $\tau$ and $n$ depend on the scale of $G$, which will be specified when specific experiments are discussed. We implemented an algorithm similar to the one in~\cite{han2013mining} to find all neighborhood patterns in $S$ from $G$.

For $SimLW$ and $SimKN$, we set parameters $\lambda_i$'s so that shorter labeled walks and $k$-hops neighbors of smaller value of $k$ have heavier weights.  Sepecifically, we set $\lambda_i=\frac{\alpha^i}{Z}$, where $0< \alpha <1$ is a constant and the normalized term $Z=\sum_{i=1}^{r}{\alpha^i}$. In our experiments, we always set $\alpha=0.5$.

In our experiments, we normalized all similarities proposed because similarities after normalization have many desirable properties like automorphism recognition. Specificaly, we normalized $SimMCNP$ and $SimNP$ to $NSimMCNP$ and $NSimNP$ using Equation~\ref{eq:NSim} so that their ranges are in $[0, 1]$, because $SimMCNP$ cannot be transformed to a distance metric, and the number of features in $SimNP$ are normally very large so that the generalized Jaccard similarity is more suitable than a distance metric.  Moreover, we normalized $SimLW$ and $SimKN$ to metrics $DSimLW$ and $DSimKN$ using Equation~\ref{eq:metric}.


\subsection{Case Studies Using Synthetic Data}\label{sec:case-study}

We generated a synthetic graph with 64 nodes $\{u_0,u_1,\ldots,u_{63}\}$ and 115 edges using the Kronecker graph model~\cite{leskovec2010kronecker}. In this synthetic graph, there are $4$ labels and each label is applied on $16$ nodes.

For each node $u$ in the synthetic graph, we searched the top-5 nodes that are most similar to $u$. For $SimNP$, we set $\tau=3$ and $n=5$.

\begin{figure}[t]
\centering
\epsfig{file=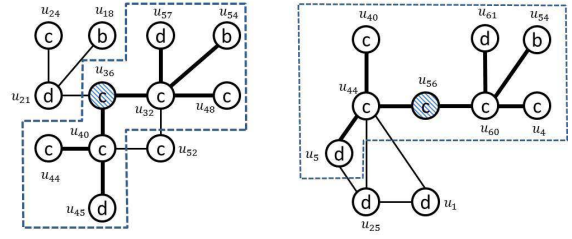, width=75mm}
\caption{The neighborhoods of nodes $u_{36}$ and $u_{56}$.}\label{fig:36-32-56}
\end{figure}

\begin{figure*}[t]
\centering
\epsfig{file=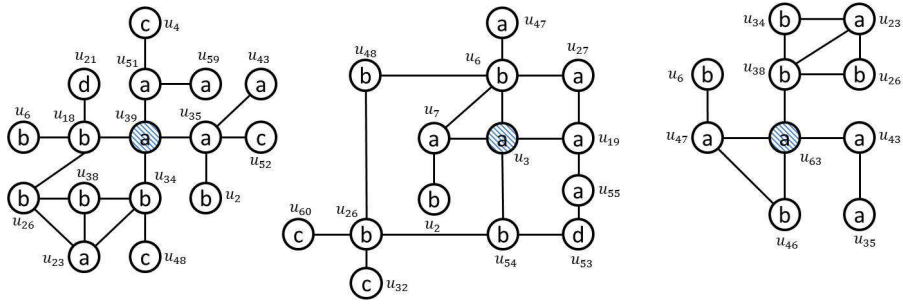, width=120mm}
\caption{The neighborhoods of $u_{39}$, $u_{3}$ and $u_{63}$}\label{fig:39-3-63}
\end{figure*}

It is not surprising that similarities based on different structural features lead to different search results. To observe the differences among similarity measures, we report some cases of similar node pairs returned by different similarities. We show neighborhoods of similar node pairs in Figures~\ref{fig:36-32-56}-\ref{fig:0-28-48}, all center nodes are marked in shadow. In those figures, we use letters $a$, $b$, $c$ and $d$ to represent labels.

\nop{
\begin{figure}[t]
\centering
\epsfig{file=42-62.eps, width=35mm}
\caption{Neighborhoods of nodes $u_{42}$ and $u_{62}$}\label{fig:42-62}
\end{figure}

Figure~\ref{fig:42-62} shows the neighborhoods of nodes $u_{42}$ and $u_{62}$. Nodes $u_{42}$ and $u_{62}$ are automorphic equivalent.  $NSimMCNP(u_{42},u_{62})=NSimNP(u_{42},u_{62})=1$. $DSimLW(u_{42},u_{62})=DSimKN(u_{42},u_{62})=0$.
}

Figure~\ref{fig:36-32-56} shows the neighborhood of node $u_{36}$. All similarities identify that node $u_{56}$ is the most similar one to $u_{36}$, whose neighborhood is also shown in the figure. Comparing these two neighborhoods, they share a big neighborhood pattern of $7$ edges, highlighted by the dashed polygons in the figure.  Moreover, the distributions of the labels over the neighborhoods are also similar.  Please note that the similarity between $u_{36}$ and $u_{56}$ cannot be found using the similarity measures based on proximity or relative proximity.


Figure~\ref{fig:39-3-63} shows another interesting example. $SimMCNP$, $SimLW$ and $SimKN$ all indicate that $u_{3}$ is the most similar node to $u_{39}$. At the same time, $SimNP$ picks $u_{63}$ as the most similar node to $u_{39}$.  Figure~\ref{fig:39-3-63} shows the neighborhoods of those nodes.  Both $u_{39}$ and $u_{63}$ are hubs in their neighborhoods, but $u_{3}$ is not a hub.  $SimNP$ can identify similarity between hub nodes because it uses neighborhood patterns frequent in the graph.  Hub nodes likely share many patterns in their neighborhoods.  $SimLW$ and $SimKN$ do not use any features that distinguish hub nodes from other nodes, and thus are not sensitive to similarity between hub nodes.  $SimMCNP$ considers the largest common neighborhood pattern.  It happens that $u_{39}$ and $u_3$ share a large common neighborhood pattern of $10$ edges in their neighborhood.  That also affect the feature vectors using labeled walks and $k$-hops nodes in $SimLW$ and $SimKN$.


\begin{figure}[t]
\centering
\epsfig{file=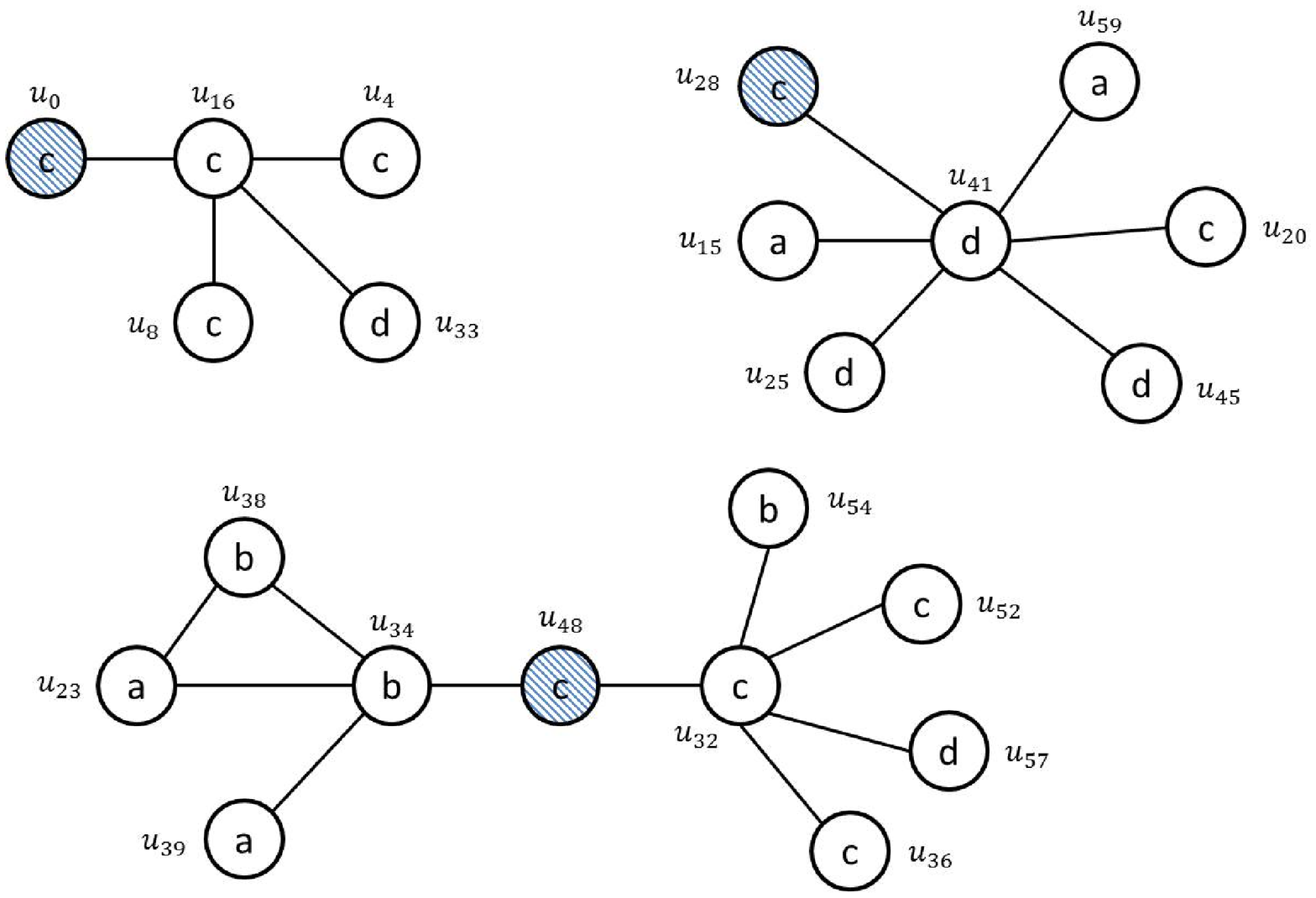, width=85mm}
\caption{Neighborhoods of $u_{0}$, $u_{28}$ and $u_{48}$.}\label{fig:0-28-48}
\end{figure}

The last example illustrates the difference between $SimLW$ and $SimKN$.
Figure~\ref{fig:0-28-48} shows the neighborhoods of nodes $u_{0}$, $u_{28}$ and $u_{48}$. According to $SimLW$, the most similar node to $u_{0}$ is $u_{48}$. According to $SimKN$, the one most similar to $u_0$ is $u_{28}$.  From  $SimLW$'s perspective, $u_{0}$ and $u_{28}$ are not similar at all because there are no common labeled walks starting from $u_{0}$ and $u_{28}$. For $SimKN$, although the 1-hop neighbors of $u_{0}$ and $u_{28}$ are totally different, the 2-hops neighbors of them are similar in labels. Thus $SimKN$ indicates that $u_{0}$ and $u_{28}$ are similar.

\begin{table*}[t]
\small
\centering \caption{The top-5 similar authors.} \label{tab:author}
\begin{tabular}{*{7}{|c}|}
    \hline
    \multirow{2}*{Rank}    &   \multicolumn{2}{|c|}{Hong Cheng}  &   \multicolumn{2}{|c|}{Spiros Papadimitriou} & \multicolumn{2}{|c|}{Jure Leskovec} \\ \cline{2-7}
    &   SimNP   & SimLW/KN &   SimNP   &   SimLW/KN   &   SimNP   &   SimLW/KN \\ \hline
    1 & Xifeng Yan & Gao Cong & Xifeng Yan & George Kollios & Deepayan Chakrabarti & Glenn Fung  \\ \hline
    2 & Jimeng Sun & Alexey Pryakhin & Jimeng Sun & Michail Vlachos & Gang Wu & Inderjit S. Dhillon  \\ \hline
    3 & Jiong Yang & J\"{o}rg Sander & Jaideep Srivastava & Gautam Das & Ji-Rong Wen & Matthias Schubert  \\ \hline
    4 & Spiros Papadimitriou & Boualem Benatallah & Hong Cheng &  Jiong Yang & Sugato Basu & Dong Xin  \\ \hline
    5 & David W. Cheung & Prasan Roy & David W. Cheung & Jignesh M. Patel & Arindam Banerjee & Roberto J. Bayardo Jr. \\ \hline
\end{tabular}
\end{table*}

\begin{table*}[t]
\centering \caption{Probability adjacency matrix of $G_1$.} \label{tab:PPAM}
\begin{tabular}{|c|c|c|c|c|}
    \hline
                   &   Professor(0)   &   Graduate(1)   &   Researcher(2) &   Engineer(3)  \\  \hline
    Professor(0)   &     0.9          &    0.25         &    0.15         &     0.1        \\  \hline
    Graduate(1)    &     0.25         &    0.9          &    0.15         &     0.15       \\  \hline
    Researcher(2)  &     0.15         &    0.15         &    0.9          &     0.2        \\  \hline
    Engineer(4)    &     0.1          &    0.15         &    0.2          &     0.9        \\  \hline
\end{tabular}
\end{table*}

\begin{table*}[t]
\scriptsize
\centering
\centering \caption{Anomaly nodes ranking results.} \label{tab:anomaly}
\begin{tabular}{*{10}{|c}|}\hline
        & ID & Label & Original Label & $SimNP$ & $SimNP$ (Original Group) & $SimLW$ & $SimLW$ (Original Group) & $SimKN$ & $SimKN$ (Original Group) \\ \hline
KG4(1)  & 146 & Professor  & Researcher &   10  &   45  &   2   &   21  &   6   &   22  \\
        & 43  & Graduate   & Engineer   &   6   &   28  &   6   &   19  &   6   &   12  \\
        & 165 & Researcher & Graduate   &   11  &   57  &   1   &   1   &   1   &   1   \\
        & 156 & Engineer   & Professor  &   12  &   56  &   2   &   23  &   2   &   23  \\ \hline
KG4(2)  & 38  & Professor  & Researcher &   4   &   45  &   1   &   29  &   2   &   17  \\
        & 55  & Graduate   & Engineer   &   24  &   27  &   5   &   4   &   3   &   6   \\
        & 213 & Researcher & Graduate   &   30  &   37  &   8   &   30  &   11  &   15  \\
        & 136 & Engineer   & Professor  &   10  &   35  &   4   &   15  &   6   &   20  \\ \hline
KG4(3)  & 66  & Professor  & Researcher &   3   &   2   &   9   &   16  &   11  &   20  \\
        & 167 & Graduate   & Engineer   &   14  &   52  &   8   &   24  &   8   &   14  \\
        & 153 & Researcher & Graduate   &   25  &   55  &   8   &   27  &   8   &   31  \\
        & 164 & Engineer   & Professor  &   7   &   54  &   7   &   28  &   6   &   29  \\ \hline
KG4(4)  & 134 & Professor  & Researcher &   6   &   24  &   14  &   19  &   13  &   23  \\
        & 39  & Graduate   & Engineer   &   20  &   22  &   1   &   5   &   1   &   3   \\
        & 197 & Researcher & Graduate   &   6   &   16  &   3   &   31  &   4   &   24  \\
        & 244 & Engineer   & Professor  &   5   &   38  &   5   &   25  &   5   &   19  \\ \hline
KG4(5)  & 158 & Professor  & Researcher &   30  &   29  &   6   &   14  &   11  &   12  \\
        & 183 & Graduate   & Engineer   &   23  &   53  &   1   &   1   &   2   &   2   \\
        & 189 & Researcher & Graduate   &   34  &   53  &   8   &   26  &   9   &   24  \\
        & 16  & Engineer   & Professor  &   1   &   2   &   3   &   27  &   1   &   22  \\ \hline
\multicolumn{4}{|c|}{\# High Rank Cases (in Top-10)}   &   10  &   2   &   19  &   4   &   16  &   4   \\ \hline
\multicolumn{4}{|c|}{\# Low Rank Cases (not in Top-20)}&   6   &   17  &   0   &   11  &   0   &   8   \\ \hline
\multicolumn{4}{|c|}{Average Rank Promotion}    &   22.45  &   &   14.15   &   &   11.15   &   \\ \hline
\end{tabular}
\end{table*}

The above examples show the common properties and differences of our proposed similarities. The results indicate that those similarities are reasonable to some extent from their own perspectives, and capture different types of similarity. An application has to pick the measure fitting the meaning of similarity in the application the best.

\subsection{Similar Authors Search}

We conducted similar authors search in the DBLP data set released by Sun~\textit{et~al.}~\cite{sun2009ranking} using the similarities we proposed. This DBLP data set contains 28,702 authors, 28,569 papers and 20 conferences from the areas of databases, data mining, machine learning and information retrieval. There are only two types of edges in this network, edges between authors and their papers, and edges between papers and conferences where papers were published. We treat this network as a labeled graph with 22 labels, which are ``author'', ``paper'', and the identities of the 20 conferences. Note that each conference only corresponds to one node, because we do not distinguish the conference held in different years.

This DBLP data set contains data up to 2009, which means all papers in the data set were published before 2009. We chose 3 young researchers who graduated around the year of 2009 as query authors. They are Hong Cheng, Spiros Papadimitriou, and Jure Leskovec. The reason we do not chose well established researchers like Christos Faloutsos, Jiawei Han and Jennifer Widom is that they are very famous and many methods have been proposed to find authors similar to them.

We used our proposed similarities to find the top-5 similar authors to the 3 young researchers. 
We do not report the results of $SimMCNP$, since it cannot handle large neighborhoods in a reasonable amount of time.
For $SimNP$, we set $\tau=100$ and $n=5$.
Table~\ref{tab:author} shows the results. Due to the structure of this DBLP data set and $r=2$, labeled walks features and $k$-hops neighbors features happen to be identical.  Consequently, the results of $SimLW$ and $SimKN$ are identical.  We report them in the columns of $SimLW/KN$.

From Table~\ref{tab:author} we find that the results of $SimNP$ and those of $SimLW/KN$ are quite different.  They capture similarity in different senses. This observation echoes the motivation of this paper -- similarity has different meanings for different people in different scenarios.

$SimNP$ identifies some authors who graduated from the same group or very similar groups as the query author. For instances, Hong Cheng and Xifeng Yan were both from Professor Jiawei Han's group, Jure Leskovec and Deepayan Chakrabarti were both Professor Christos Faloutsos's students. Athough Spiros Papadimitriou and Xifeng Yan were not from the same group, the two groups (Christos Faloutsos's group and Jiawei Han's group) they did their Ph.D.\ studies in are similar in terms of publication venues and collaborators. The neighborhood patterns in this DBLP data set reflect some patterns of an author's publication style, such as publishing with a certain number of co-authors in a certain conference and co-authoring a certain number of papers with a specific researcher.  For young researchers, especially when they were still Ph.D.\ students, such publication patterns were often deeply influenced by their advisors. Thus, from the publication pattern perspective, the most similar authors are likely those graduated from the same group or very similar groups.

$SimLW/KN$ captures another interesting meaning of similarity.  These measures return authors who share similar publication venues with the query authors. Due to the structure of this DBLP network data set, the labeled walks of length $1$ do not have much impact on the similarity score -- there is only one type of labeled walk of length $1$, ``author-paper''. Labeled walks from authors to conferences are the features playing the major role in $SimLW/KN$.

\subsection{Anomaly Nodes Ranking}

We generated a labeled graph using the Kronecker graph model~\cite{leskovec2010kronecker}.  Then, we added some abnormal nodes to it. Our task is to check whether those proposed similarity measures can help to find those abnormal nodes.  Specifically, the initiator graph $G_1$ has 4 nodes with labels ``Professor'', ``Graduate'', ``Researcher'' and ``Engineer'', respectively. The probability adjacency matrix $(P_1)_{4 \times 4}$ is shown in Table~\ref{tab:PPAM}.

Here ``Professor(0)'' denotes that the 0-th node in $G_1$ is with label Professor. Using this probability adjacency matrix, we applied the Kronecker graph generator~\cite{leskovec2010kronecker} to generate the probability adjacency matrix $P$ of $KG4$, which is a $256 \times 256$ matrix, because in $KG4$ there are $4^4=256$ nodes. The entry in $P$ follows
\begin{equation}\label{eq:kronecker}
    P(u,v) = \Pi_{i=1}^{4}{P_1(\lfloor \frac{u}{4^{i-1}} \rfloor (mod \ 4), \lfloor \frac{v}{4^{i-1}} \rfloor (mod \ 4))}
\end{equation}
The label of node $u$ is $L(u)=L(u(mod 4))$, where $L(0)$=Professor, $L(1)$=Graduate, $L(2)$=Researcher and $L(3)$=Engineer as in the prototype graph $G_1$. $KG4$ was generated according to $P$. We preserved every potential edge $(u,v)$ with probability $P(u,v)$ independently.

From Equation~\ref{eq:kronecker}, one can see that a node is more likely linked to nodes that have the same label. We add some abnormal nodes that violate this tendency. We randomly picked 4 nodes in KG4, one for each label, and switch their labels. Specifically, for a node picked randomly, if its label is ``Professor'', we changed it to ``Engineer''; if its label is ``Graduate'', we changed it to ``Researcher''; if its label is ``Researcher'', we changed it to ``Professor''; and if its label is ``Engineer'', we changed it to ``Graduate''. From the probability adjacency matrix of $G_1$ in Tabel~\ref{tab:PPAM} and Equation~\ref{eq:kronecker}, we can see that after switching labels, those picked nodes may look ``weird'' based on their connections in the network. Thus, we call these 4 nodes \textbf{planted anomaly nodes}.

We designed a very simple algorithm which ranks nodes according to how abnormal they look like. Specifically, we first partitioned nodes into 4 groups according to their labels. Note that for each group we have 64 nodes. Then, for each node $u$, we computed the top-5 nodes that are most similar to $u$ from the group $u$ belongs to. We calculated the sum of similarity scores between $u$ and its top-5 similar nodes, and used this total to measure how ``normal'' $u$ is. We call this total the \textbf{normality index} of $u$.  For $SimLW$ and $SimKN$, since we used the distance metrics $DSimLW$ and $DSimKN$, the normality index of $u$ is the negated sum of the distances between $u$ and $u$'s top-5 closest nodes. Last, for each group, we sorted all nodes in the group in the normality index ascending order. Those top-ranked nodes in every group are regarded as most abnormal. Clearly, the higher those 4 planted anomaly nodes are ranked, the better our methods are.

We randomly generated 5 data sets, denoted by $KG4(1)$-$KG4(5)$, and we used $SimNP$, $SimLW$ and $SimKN$ to rank nodes. For $SimNP$, we set $\tau=3$ and $n=5$. 

Table~\ref{tab:anomaly} reports the results of anomaly nodes ranking, where column ``Label'' lists the labels of the planted anomaly nodes after label switching, column ``Original Label'' lists the original labels of the planted anomaly nodes. For each planted anomaly node, we report two ranks, the rank without switching the label and the one with switching the label.  It is expected that by switching the label, the change of the rank is significant.  The columns $SimNP$, $SimLW$ and $SimKN$ list the ranks of the planted anomaly nodes with switching the labels, while the columns ``$SimNP$ (Orignial Group)'', ``$SimLW$ (Orignial Group)'', and ``$SimKN$ (Orignial Group)'' list the ranks of the planted anomaly nodes dwithout switching the labels.   The row ``Average Rank Promotion'' provides the average number of positions elevated for a planted anomaly node by switching the label. Please node that some planted nodes are ranked high and some are ranked low even without switching the labels.  Therefore, we believe that our results are representative.

From Table~\ref{tab:anomaly}, we observe that $SimLW$ ranks the planted anomaly nodes high (in top-$10$) most of the cases ($19$ of the $20$ cases in $5$ trials).  To this extent, $SimLW$ is capable of catching the planted anomaly nodes, and likely is capable of catching this type of anomaly nodes in general.  $SimNP$ does not rank those planted anomaly nodes high.  At the same time, as expected, all the three proposed similarity measures promote the anomaly ranks of the planted anomaly nodes substantially, as indicated by the row ``Average rank promotion''.

Although the neighborhood patterns of nodes in the randomly generated graphs are unknown, according to Equation~\ref{eq:kronecker} and $P_1$ (Table~\ref{tab:PPAM}), the model generating the graphs encourages nodes to link to nodes with the same labels. The neighbors of nodes in the graphs tend to have some underlying patterns. In $SimLW$, the labeled walks of length $1$ as features capture neighbors, and so do $1$-hop neighbors in $SimKN$. Therefore, these two methods are capable in this anomaly nodes ranking task.

\subsection{Nodes Clustering}

The last experiment conducted is nodes clustering in social networks. We adopted the netscience co-author network~\cite{newman2006finding} as our data set, which contains 1589 authors and 2743 edges. One may note that this network is unlabeled, but unlabeled graphs can be seen as labeled graphs with only one type of label on all nodes. We can still run our proposed methods on the netscience co-author network. Since all nodes are with the same label, labeled walks and $k$-hops neighbors degenerate into counting $k$-hops degree of nodes. Thus, in this experiment, we only use neighborhood patterns as features for clustering nodes.

We computed the normalized similarity (Theorem~\ref{thm:jaccard}) between every pair of nodes in the network using $SimNP$. To decide the feature set $S$ of neighborhood patterns, we set $\tau=10$ and $n=5$. Then, we built a node-node similarity matrix and ran a simple spectral clustering algorithm~\cite{ng2002spectral} by setting the number of clusters to 5. Figure~\ref{fig:cluster} shows the result on the largest connected component of the netscience co-author network, which has 379 nodes.  We use different colors to indicate nodes in different clusters.

\begin{figure}[t]
\centering
\epsfig{file=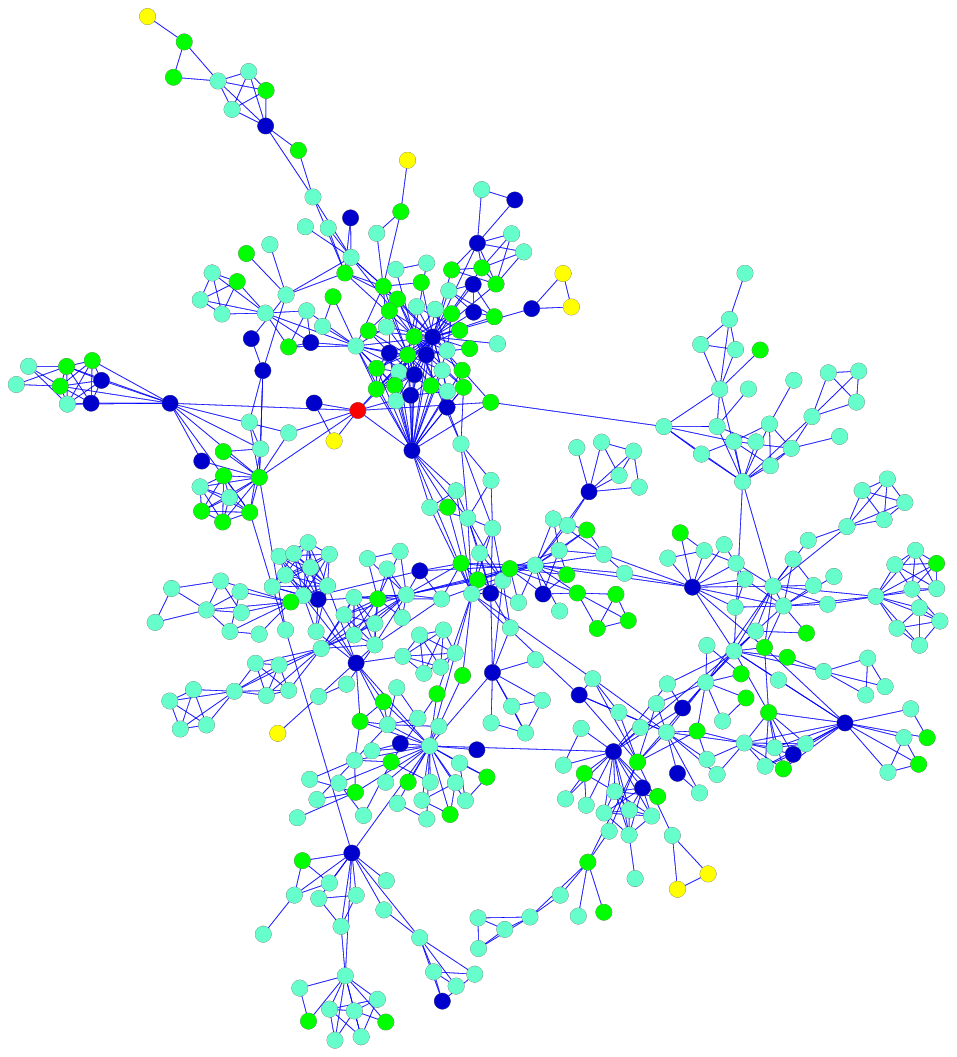, width=5in}
\caption{Nodes clustering result on the largest connected component of the NetScience co-author network.}
\label{fig:cluster}
\end{figure}

Using $SimNP$, we partitioned nodes into groups according to their structural roles in the network. Most of the nodes in the dark-blue cluster are hub-like nodes, most of the nodes belonging to the yellow cluster are isolated nodes (outliers), most of the nodes in the green cluster are less isolated nodes, and most of the nodes in the light-blue cluster are main-stream nodes. Only one node is assigned to the red cluster, which is a hub-like node.  The results here are similar to the results in~\cite{henderson2012rolx}. Please note that, unlike the traditional hub/outlier detection algorithms on graphs, our method did not use the whole topological information of the graph, but only used similarity between pairs of nodes. Unlike~\cite{henderson2012rolx}, we did not conduct feature selection/engineering or model selection.  We just applied a simple spectral clustering algorithm on the node-node similarity matrix. This set of experimental results demonstrate that the node similarity based on neighborhood patterns is promising for distinguishing nodes with different structural roles.

\section{Conclusions}\label{sec:con}

In this paper, we systematically investigated the problem of measuring in-network node similarity based on neighborhoods. Neighborhood based node similarities can capture various meanings of similarity that are different from most of the existing proximity or relative proximity based similarities, such as SimRank and Guilt-by-Association. We proposed a unified parametric model for neighborhood based similarity, which is flexible for plugging in different features and assembling functions to obtain similarities with different meanings. At the same time, the model remains simple.  Four different similarities based on different features were developed and proved to be instances of our unified parametric model. We explored desirable properties of our proposed similarities, such as how to transform them into distance metrics and normalization. We also discussed the computational costs of different similarities and analyzed interesting tradeoffs between topological matching and computational efficiency. Last, extensive empirical studies on both synthetic data and real-world data were conducted, and the results demonstrated the effectiveness of our proposed similarities.

This paper is the beginning of an exciting journey on developing in-network similarity measures and applications.  There are a series interesting and important problems for future work.  For example, it is interesting to explore efficient and scalable algorithms to compute and approximate the similarity measures over large graphs.  Moreover, it is useful to further refine the spectrum of similarity measures representing more tradeoffs between computational cost and strictness of neighborhood matching and addressing meaningful application needs.  Systematic applications of the proposed model and similarity measures in social network mining applications will also be exciting.

\bibliographystyle{abbrv}
\bibliography{nodes_similarity}
\end{sloppy}
\end{document}